\documentclass[twocolumn]{aa}  

\usepackage{graphicx}
\usepackage{gensymb}
\usepackage[hidelinks]{hyperref}
\usepackage{color}
\usepackage[normalem]{ulem}
\usepackage{subcaption}
\usepackage{caption}
\usepackage{xcolor}

\usepackage{txfonts}
\begin{document}

   \title{Radio jet precession in M81*}

   \author{S.D. von Fellenberg \inst{1}
          \and
          M. Janssen \inst{1}
          \and
          J. Davelaar \inst{3,4}
          \and
          M. Zaja\v{c}ek \inst{5}
          \and
          S. Britzen\inst{1}
          \and
          H. Falcke \inst{2}
          \and 
          E. K{\"o}rding \inst{2}
          \and 
          E. Ros \inst{1}
          }

   \institute{Max Planck Insititute for Radio Astronomy, Auf dem H{\"u}gel 69, Bonn D-53121, Germany
              \email{sfellenberg@mpifr-bonn.mpg.de}
         \and Department of Astrophysics, Institute for Mathematics, Astrophysics and Particle Physics (IMAPP), Radboud University, PO Box 9010, 6500 GL Nijmegen, the Netherlands
        \and Department of Astronomy and Columbia Astrophysics Laboratory, Columbia University, 550 W 120th Street, New York, NY 10027, USA
        \and Center for Computational Astrophysics, Flatiron Institute, 162 Fifth Avenue, New York, NY 10010, USA 
        \and Department of Theoretical physics and Astrophysics, Faculty of Science, Masaryk University, Kotl{\'a}{\v r}sk{\'a} 2, 611 37 Brno, Czech Republic}
   \date{Received September 15, 1996; accepted March 16, 1997}
 
  \abstract{
  We report four novel position angle measurements of the core region of M81* at 5GHz and 8GHz, which confirm the presence of sinusoidal jet precession of the M81 jet region as suggested by \cite{Marti-Vidal2011}. The model makes three testable predictions on the evolution of the jet precession, which we test in our data with observations in 2017, 2018, and 2019. Our data confirms a precession period of $\sim7~\mathrm{yr}$ on top of a small linear drift. We further show that two 8 GHz observation are consistent with a precession period of $\sim 7~\mathrm{yr}$, but show a different time-lag w.r.t. to the 5 GHz and 1.7 GHz observations. We do not find a periodic modulation of the light curve with the jet precession, and therefore rule out a Doppler nature of the historic 1998-2002 flare. Our observations are consistent with either a binary black hole origin of the precession or the Lense-Thirring effect.}

   \maketitle
%

\section{Introduction}
The galaxy M81 appears as a bright radio source and is located at a distance of $3.36\pm0.34~\mathrm{Mpc}$
\citep{Freedman1994}.
The black hole in the center of M81 belongs to the class of the low-luminosity AGN (LLAGN) and exhibits relatively weak radio emission ($F_{\nu=4.8\mathrm{GHz}} \sim 150~\mathrm{mJy}$, e.g.: \cite{Brunthaler2006}). It is the closest LLAGN to Earth. Due to its high apparent luminosity, it serves as an optimal test-bed to characterize this class of accreting black holes \citep{Markoff2008}. As such it may be the best candidate to bridge the accretion processes of LLAGN and to that of Sgr~A* \citep{GRAVITY_flux2020}. 
M81 shows slow changes in radio flux density on yearly time scales \citep[e.g.][]{Ho1999}. Further, it shows fast intra-day flare-like variability at mm-wavelengths \citep{Sakamoto2001}. This radio and mm behavior is consistent with van der Laan expanding-blob-scenario \citep{vanderLaan1966}. Here, the mm-variability is created by blobs that expand as they move along the jet, where they become observable in the radio  \citep{Ho1999, Sakamoto2001}.   
In the X-ray, M81 is detected with luminosity of $\sim10^{40}~\mathrm{erg/s}$, and flux changes on the order of a few ten percent \citep{Ishisaki1996}. 
The overall SED was modeled with a jet-dominated ADAF model by \cite{Markoff2008} using a set of simultaneous multi-wavelength observations of M81 and shows remarkable similarity to that of Sgr~A* \citep[e.g.,][]{vonFellenberg2018}. 

M81*, the core region of M81 is a regular target for global VLBI observations, in particular since the explosion of the radio-luminous supernova SN 1993J \citep{Ripero1993, Weiler2007}. This supernova was located at a close angular separation to M81*, and was thus frequently used as a calibration source. 
In VLBI observations, M81* is typically marginally resolved and shows a jet in north-eastern direction. 
\cite{Bietenholz2000} and \cite{Bietenholz2004} determined that M81 is at rest with respect to SN1993J. Further, they found a frequency-dependent shift of peak-brightness of M81*. This is consistent with the known core-shift-effect of many AGN jets \citep{Marcaide1984, Lobanov1998, Kovalev2008}, which is believed to be caused by synchrotron self-absorption of photons in the jet plasma. In this picture, the apparent shift of the luminous component results from the increasing opacity (and thus increasing luminosity) as function of wavelength \citep[][]{Blandford1979, Konigl1981, Falcke1995, Marscher1996, Davelaar2018}. 
M81* shows a decreasing core-size ($\Theta$) as function of frequency, with an almost linear relationship: $\Theta\propto \lambda^{\sim 0.9}$ \citep{Bartel1982, Kellermann1976, Bietenholz2000, Markoff2008}. This relationship holds down to mm-frequencies, with a confirmation at 43 GHz \citep{Ros2012}, and one at 87 GHz \citep{Jiang2018}, which report a relation of $\Theta\propto \lambda^{0.89\pm0.03}$.

\cite{Marti-Vidal2011} confirmed the basic findings reported in earlier works. They further showed that the M81* intensity peak is shifted as function of frequency along the direction of the jet using VLBI measurements from 1993 to 2005. Its size increases with decreasing frequency. The authors determined the location of the jet base to within $20~\mathrm{\mu as}$ of the black hole, and constrained the black hole mass to $\sim2\times 10^7~\mathrm{M_\sun}$ using the strongly-magnetized accretion flow scenario \citep{Kardashev1995}. \cite{Alberdi2013} extend the temporal baseline of observation to the year 2012 and confirmed the basic findings reported in \cite{Marti-Vidal2011}. By fitting elliptical Gaussian models to the central source and, if detected, the jet component, they derived a sinusoidal modulation of the jet position angle. They found a precession period of $\sim 7~\mathrm{yr}$ on top of a linear increase of the position angle by $\sim 0.5 \mathrm{\degree / yr}$. They found this precession to be present both in their 5 GHz observations, as well as in their 1.7 GHz observations. However, they found a lag of $1.9\pm0.4$ years for the precession between the frequency bands, which they interpreted as a core-shift effect. In this picture, the jet shows a differential cork-screw-like precession and different frequencies probe different regions along the jet. 
Lastly, they connected the observed jet precession with a four year flare, and argued that the increase in flux is caused by Doppler boosting of the jet along the line of sight. 

Their model therefore provides three testable hypotheses:
\begin{enumerate}
    \item A prediction of the position angle of the core component as function of time;
    \item A prediction of this modulation at different frequencies;
    \item A prediction of the expected flux level at a given time point.
\end{enumerate}
In this letter, we investigate whether the three predictions hold against new VLBI observations of M81* obtained in the years 2017, 2018, and 2019. 

\section{Observations and data reduction}
In total we analyze four sets of observations of M81*. The first set, obtained by the European VLBI Network (EVN) at 5 GHz in June 2017, was a dedicated observation to measure the position angle of M81* (Prog. ID ED042, PI Jordy Davelaar). The second set consists of an observation at 8 GHz obtained in June 2018 at the Very Long Baseline Array (VLBA) (Prog. ID BJ090, PI Wu Jiang), with the intent to obtain phase-referenced observations of the core-shift effect in M81*. Lastly, we used 5 GHz and 8 GHz VLBA observations obtained in 2019 dedicated to studying the jet-components in M81* (Prog. ID BJ099, PI Wu Jiang).
All data was reduced using the rPicard\footnote{\url{https://bitbucket.org/M_Janssen/picard}.} VLBI pipeline version v7.1.5 \citep{Janssen2019A}, which makes use of CASA v6.5 \citep{2022arXiv221002276T} and the latest VLBI features \citep{2022arXiv221002275V}\footnote{For reproducibility, the pipeline parameters are uploaded to \url{https://doi.org/10.5281/zenodo.7642852}}.
To derive the position angle, we fit all data with a single Gaussian model using Difmap \citep{Shepherd1997}. In all cases, the source is marginally resolved; however, we opted for a simple description by a singular Gaussian component to derive a robust  measurement of the position angle. \autoref{tab:posangle} reports the dates, frequency bands, derived values, and the respective imaging $\chi_{\rm{reduced}}^2$ values. The corresponding maps and models are shown in \autoref{appendix:models}.

In order to obtain an as complete picture of M81* as possible, we have searched the VLBA and EVN archives for available observations since 2012. Several observations at higher frequencies exist, which we do not study in this letter (e.g., BJ086, BB303). Three more observations in L, S, or X-band exist: RP023A (EVN), RP023B (EVN), and BD185 (VLBA). Those, unfortunately, do not allow a determination of the position angle as the data quality is not sufficient. In RP023A, no long baseline stations participated in the observations. In  RP023B, several telescopes suffered from sensitivity losses at the start of the scan, possibly due to being late on source. When the bad measurements are flagged, the scan durations were no longer long enough to obtain good fringe solutions. In BD185, we found large instrumental delay corrections of $\sim$100\,ns from the calibrator sources and only a few robust fringe detections on M81 over the full Nyquist search window. These issues possibly originate from an error in the clock search at the correlator.

We have further validated that we can reproduce the values published in \cite{Alberdi2013} with one example (BB293 B), which gives consistent results.
\begin{table}[]
    \centering
    \begin{tabular}{c|c|c|c}
         Date & Band& Position Angle $[\degree]$  & $\chi_{\rm{red}}^2$\\
         \hline
         \hline
         2017-06-21& C & 72.06 & 2.3\\
         2018-06-10& X & 81.06 & 1.9\\
         2019-11-04& C & 81.69 & 4.9\\         
         2019-11-04& X & 73.00 & 3.8\\  
         \hline
    \end{tabular}
    \caption{Date, observation band, derived position angle and reduced $\chi^2$ of the observation analysed in this study.}
    \label{tab:posangle}
\end{table}

\subsection{EVN Observation}
We analyze a set of 5 GHz VLBI observations carried out by the EVN observatory, which were obtained on the 21st of June, 2017 and which lasted for about $12~\mathrm{h}$. The IR, YS, TR, SH, NT, MC, WB, JB, and EF\footnote{Full name and location given in \autoref{tab:participating_stations}} stations participated in the observations, and apart from a full loss of the RR polarization of the EF station, no major technical difficulties occurred.
The data were calibrated using observations of J0958+6533. 

\subsection{VLBA BJ090 observation}
We analyze the 8 GHz sub-set of the observations carried out in the VLBA BJ090 observation campaign from the 10th of June, 2018. The data included the BR, FD, HN, KP, LA, MK, NL, OV, PT and SC stations and used OJ287, J0954+658 and J1331+305 as calibration sources. No major technical difficulties occurred during the observations, with overall good data quality.  
The data was reduced in the same fashion as the EVN observations. Apart from 8 GHz observations (X-band), the BJ090 observations included K-, Q-, and W-band observations which we did not include in the analysis.

\subsection{VLBA BJ099 observation}
We analyze the 5 GHz and 8 GHz sub-sets of the observations carried out in the VLBA BJ099 observation campaign, which was carried out on the 4th of November in 2019. The data included the BR, FD, HN, KP, LA, MK, NL, OV, PT and SC stations and used OJ287 and J0954+658 as calibration sources. While the 5 GHz observations showed overall agreeable data quality, the 8 GHz data set suffered from poor observations by the PT and SC stations. In order to derive a position angle measurement in the 8GHz band, we excluded the baselines to the PT and SC stations, the latter of which contributes the longest baselines. For both observations, the fit is relatively poor, with reduced $\chi^2$ values between four and five. Apart from 5 GHz and 8 GHz observations (C-band and X-band), the BJ099 observations included K-, Q-, and W-band observations which we did not include in the analysis. 

\section{Results}
\begin{table*}[]
    \centering
    \begin{tabular}{c|c|c|c}
         Parameter & Sinusoidal model up to 2012 & Sinusoidal model up to 2019 & Linear drift model up to 2019 \\
         \hline
         \hline
         $\theta_0$& $(62.9\pm0.8)~\mathrm{\degree}$ & $(63.1\pm0.8)~\mathrm{\degree}$& $62.4\pm0.2~\mathrm{\degree}$\\
         $\beta$& $(0.4\pm0.1))~\mathrm{\degree/yr}$ & $(0.5\pm0.1)~\mathrm{\degree/yr}$& $0.2\pm0.2~\mathrm{\degree/yr}$ \\
         $A$& $(6.9\pm0.8)~\mathrm{\degree}$ & $(6.9\pm0.8)~\mathrm{\degree}$& -\\
         $T$& $(6.7\pm0.2)~\mathrm{yr}$& $(6.9\pm0.2)~\mathrm{yr}$& -\\
         $1996-t_0$& $-(3.06\pm0.24)~\mathrm{yr}$ & $(-3.15\pm0.25)~\mathrm{yr}$& -\\
         \hline
    \end{tabular}
    \caption{Best fit values of the sinusoidal and linear drift model derived from a $\chi^2$ fit to the observed data. First column shows the model parameters for the different model and data used. The second column shows the best fit values derived from fitting a sinusoidal model (\autoref{eqn:model}) to the data up to 2012 i.e., the based on data of \cite{Marti-Vidal2011, Alberdi2013}. The third column shows values derived from the sinusoidal model including the new data from 2017 and 2019. The third column shows the best values of linear drift model (\autoref{eqn:lin_model}), including the new data from 2017 and 2019.}
    \label{tab:model-fit}
\end{table*}
In the following sections we proceed by testing the three hypothesis presented in \cite{Marti-Vidal2011} and \cite{Alberdi2013}.
\subsection{Sinusoidal jet procession}\label{sec:sinus_model}
\begin{figure}
    \centering
    \includegraphics[width=0.485\textwidth]{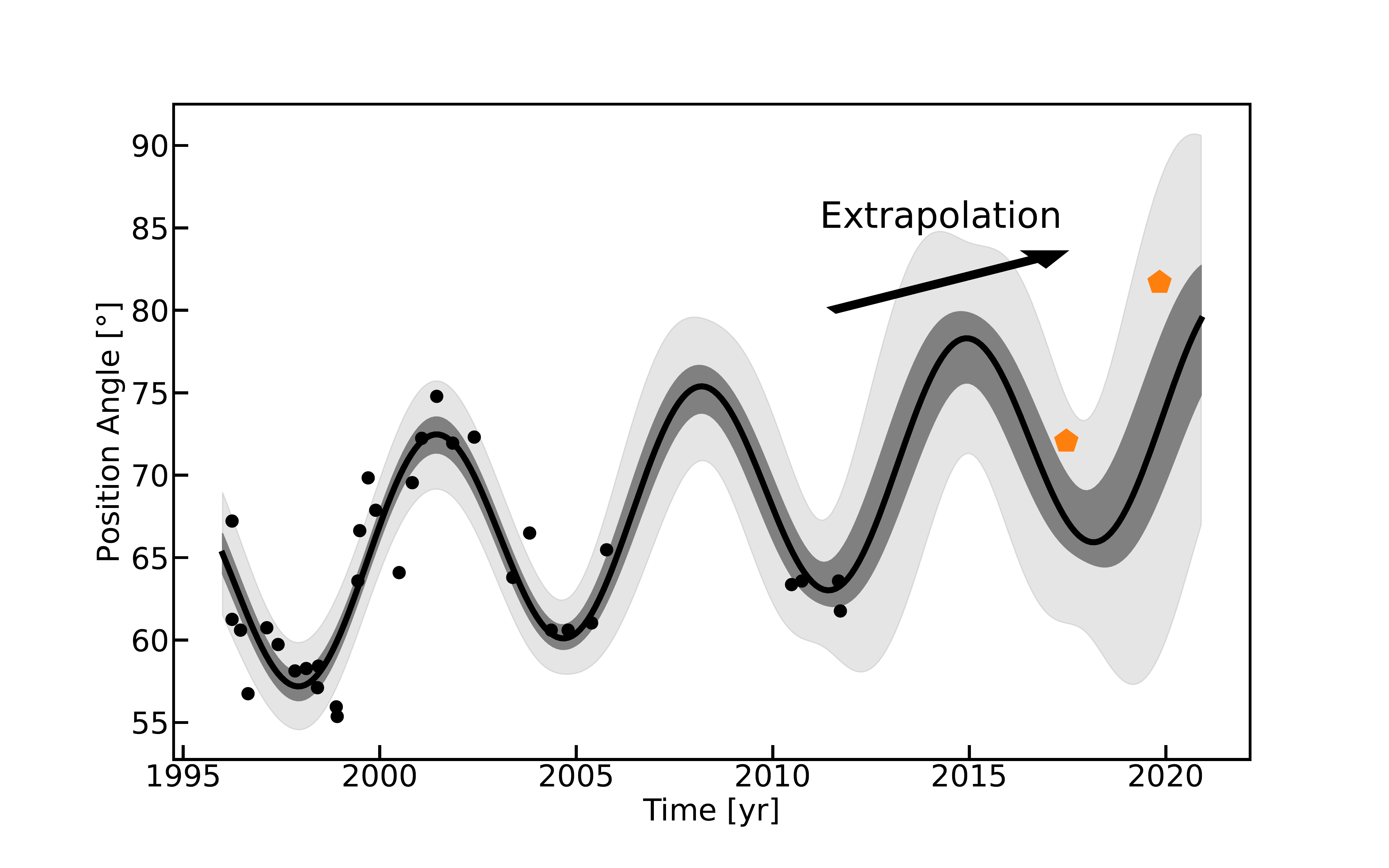}
    \caption{Black dots show the position angle measurement as function of time up to the year 2012. The orange pentagons show the 5 GHz data derived in this work. The black line shows the best fit model to the data up to 2012 by \cite{Alberdi2013}, excluding the new measurements, and extrapolated to the year 2021. The grey shaded region indicates the $1\sigma$ and the $3\sigma$ contours.}
    \label{fig:fit-2011}
\end{figure}

We adapt the model proposed in \cite{Marti-Vidal2011}:
\begin{equation}
    \theta(t) = \theta_0 + A \cdot \sin \left(\dfrac{2\pi}{T}(t-t_0)\right) + \beta\cdot(t-t_0), \label{eqn:model}
\end{equation}
where $\theta(t)$ is the position angle as function of time, $\theta_0$ is the initial position angle value at time $1996-t_0$, $T$ is the oscillation period  in years and $\beta$ is the linear drift slope in degrees per years. 
We extract the measurements presented by \cite{Marti-Vidal2011}, and fit the data up to the year 2012. The second column of \autoref{tab:model-fit} reports the values derived in this manner. \autoref{fig:fit-2011}
shows the best fit. The shaded regions show the $1\sigma$, and $3\sigma$ contours derived by sampling $1000$ model realizations from the best fit values and corresponding errors, where we assumed a Gaussian distribution and took into account the covariance of the best fit values. The model is able to predict the observations in 2017 and 2019 well, with the predicted values within the $3\sigma$ contour. \autoref{fig:fit_2017} shows the best fit model when the new data is included, the best fit parameters are reported in the third column of \autoref{tab:model-fit}. 

Unfortunately, it is not straightforward to determine the uncertainty of the position angle accuracy. We estimate the uncertainty of each data point by fitting the evolution of the data by a non-parametric Gaussian process\footnote{We follow the scikit-learn cookbook for Gaussian process regression.}. The uncertainty of the Gaussian process model is then used as a proxy of the uncertainty of a datum at a given time. This allows us to do a model selection test against a more simple baseline model: a simple linear drift (\autoref{eqn:lin_model}, \autoref{fig:linear_mod}). We report the best fit values of this model in the fourth column of \autoref{tab:model-fit}. The reduced $\chi^2$ of the linear drift model is $\chi^2_{\rm{linear}} \approx 38.5$, where as the sinusoidal model has $\chi^2_{\rm{sinusoidal}}\approx4.5$. The sinusoidal model is favored over a simple linear drift.

\subsection{Frequency dependent jet-modulation}
\begin{figure}
    \centering
    \includegraphics[width=0.485\textwidth]{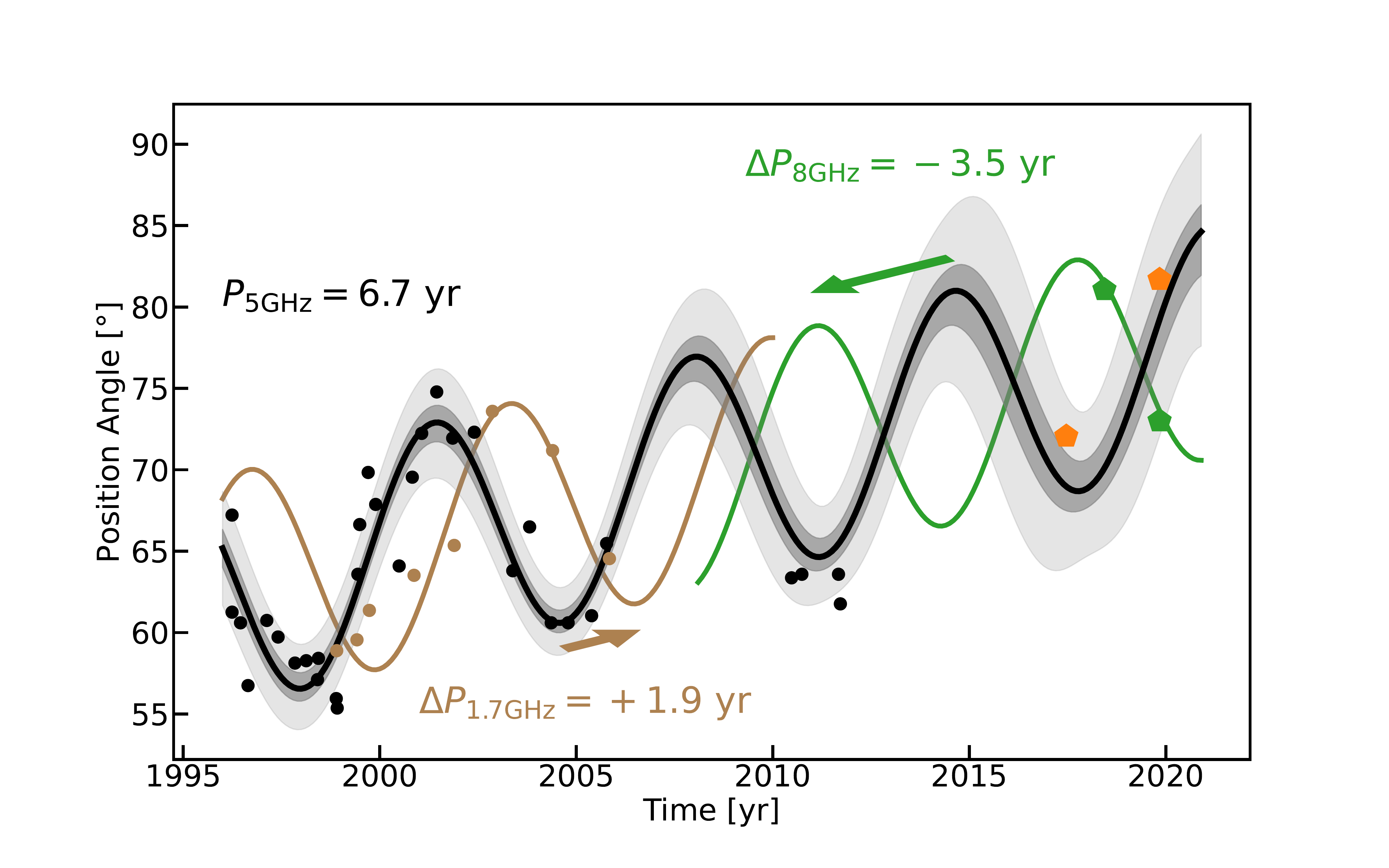}
    \caption{Multi-frequency data of the core position angle of M81*. The black points indicate the $5\rm{GHz}$ position angle as presented by \cite{Marti-Vidal2011}, the orange pentagons show the new 2018 and 2019 measurements. The black line with gray contours ($1,2,3\sigma)$ shows the best fit to all $5\rm{GHz}$ data. The brown points show the $1.7\rm{GHz}$ data; the brown line shows the best fit to the $5\rm{GHz}$ data, shifted by $1.9~\mathrm{yr}$ as suggested by \cite{Marti-Vidal2011}. The green points show the $8\rm{GHz}$ measurement; the green line shows the best fit $5\rm{GHz}$ model shifted by $\sim-3.5~\mathrm{yr}$.}
    \label{fig:frequency_dependet}
\end{figure}

\cite{Marti-Vidal2011} reported a sinusoidal modulation of the M81* core position angle both at $5~\mathrm{GHz}$ as well as at $1.7 ~\mathrm{GHz}$ and found a $(1.9\pm0.4)~\mathrm{yr}$ time lag between the frequency bands. This was interpreted as differential cork-screw-like precession where the core-shift effect \citep{Konigl1981, Marscher1996} leads to longer wavelengths probing regions at a larger separation from the black hole. The model therefore makes a prediction for our higher frequency observation at 8 GHz. Typically a linear relation ($r_c \propto \nu^{-1}$) for the core-shift is found \citep[e.g.,][]{Sokolovsky2011}, and we thus expect a similar lag of $1.9~\mathrm{yr}$ in the 8 GHz observations but in opposite direction. \autoref{fig:frequency_dependet} shows that while the observations are consistent with the period and amplitude of the oscillation, the data favor a larger time lag of $\Delta P_{\mathrm{8GHz}}\sim 3.5~\mathrm{yr}$. This result is driven by the 2019 observation, which suffered from poor data quality. If one disregards the 2019 data, the observations are consistent with a shift of $1.9$ years (see \autoref{fig:shift_1.9years}). This illustrates that the available data set is insufficient to test this hypothesis, and future observations are required for a more definitive statement.

\subsection{No precession-caused flux modulation}
\begin{figure}
    \centering
    \includegraphics[width=0.485\textwidth]{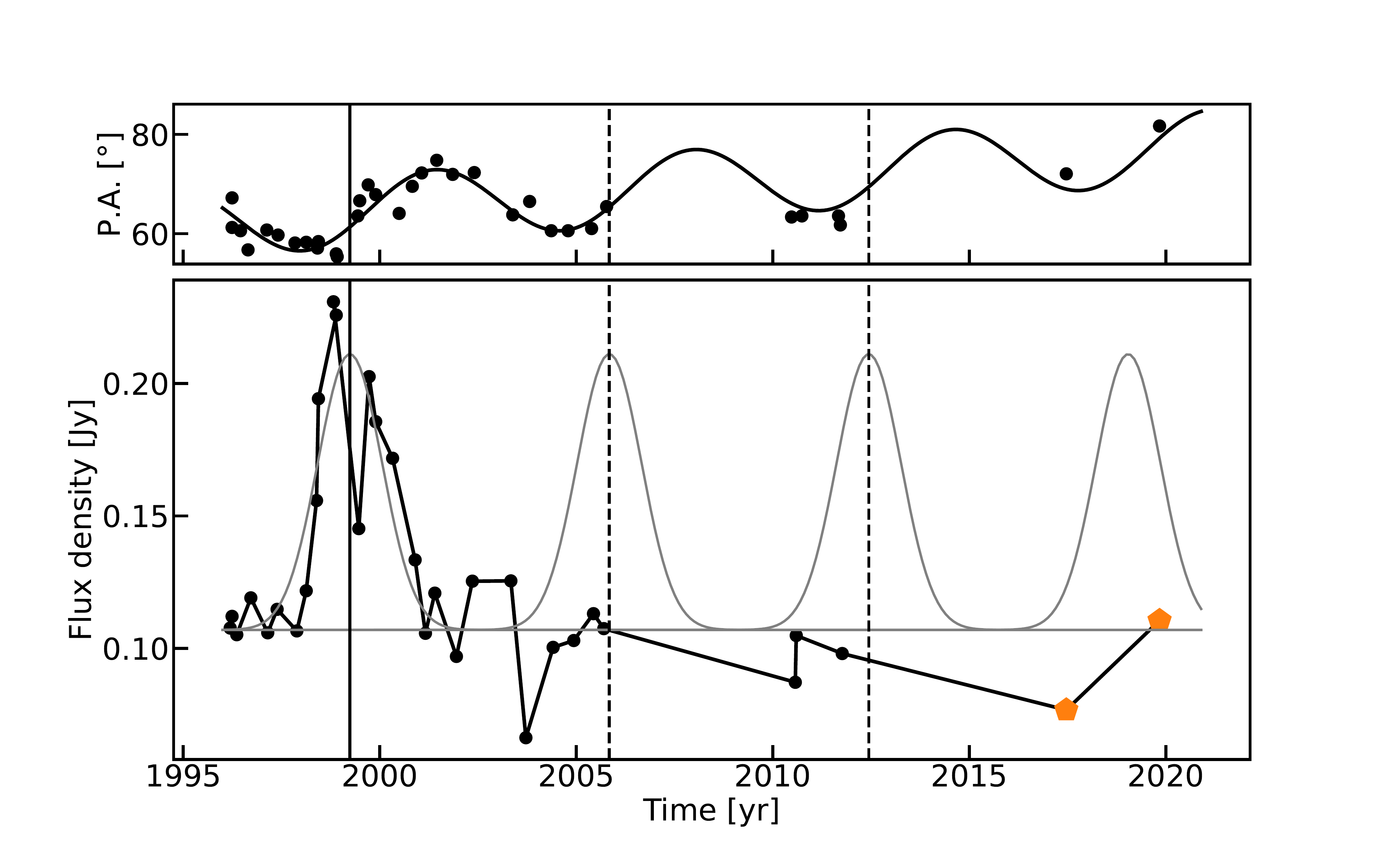}
    \caption{The top: position angle measurements alongside the best fit model (\autoref{eqn:model}). Lower panel: the 5 GHz light curve of M81*. The black lines indicate the data extracted from \cite{Marti-Vidal2011, Alberdi2013}, the orange pentagon denotes our newest measurement. The gray line shows a best fit Gaussian model which approximates the observed flare. The vertical line indicates the peak of the Gaussian flare model. The Gaussian flare shape is shifted by the precession period ($6.9~\mathrm{yr}$).}
    \label{fig:flux}
\end{figure}

\cite{Marti-Vidal2011} proposed that the flare observed in the years 1998 -- 2002 was caused by the variable Doppler boosting of the precessing jet as the viewing angle periodically changes in the direction of the observer (see also \cite{Ros2012}). The top panel of \autoref{fig:flux} shows the best fit model and the position angle measurements. We include the two newest 5 GHz flux density measurement in the bottom panel. We fit a simple Gaussian + flux offset model to the flare data. The best fit model is plotted in gray in \autoref{fig:flux}. If the flare is caused by the precession of the jet, one expects the re-occurrence of a similar flare shape shifted by the period. We illustrate this by plotting shifted realizations of the best fit Gaussian flare model, shifted by multiples of the precession period. It is clear that the low flux measurements in the years after 2002 are inconsistent with such a scenario. However, given the sparse sampling of the light curve, it is possible that the observations miss flaring activity if the correlation to the precession period is not very tight. 

\subsection{A precession-nutation model for M81*}
\begin{figure}
    \centering
    \includegraphics[width=0.485\textwidth]{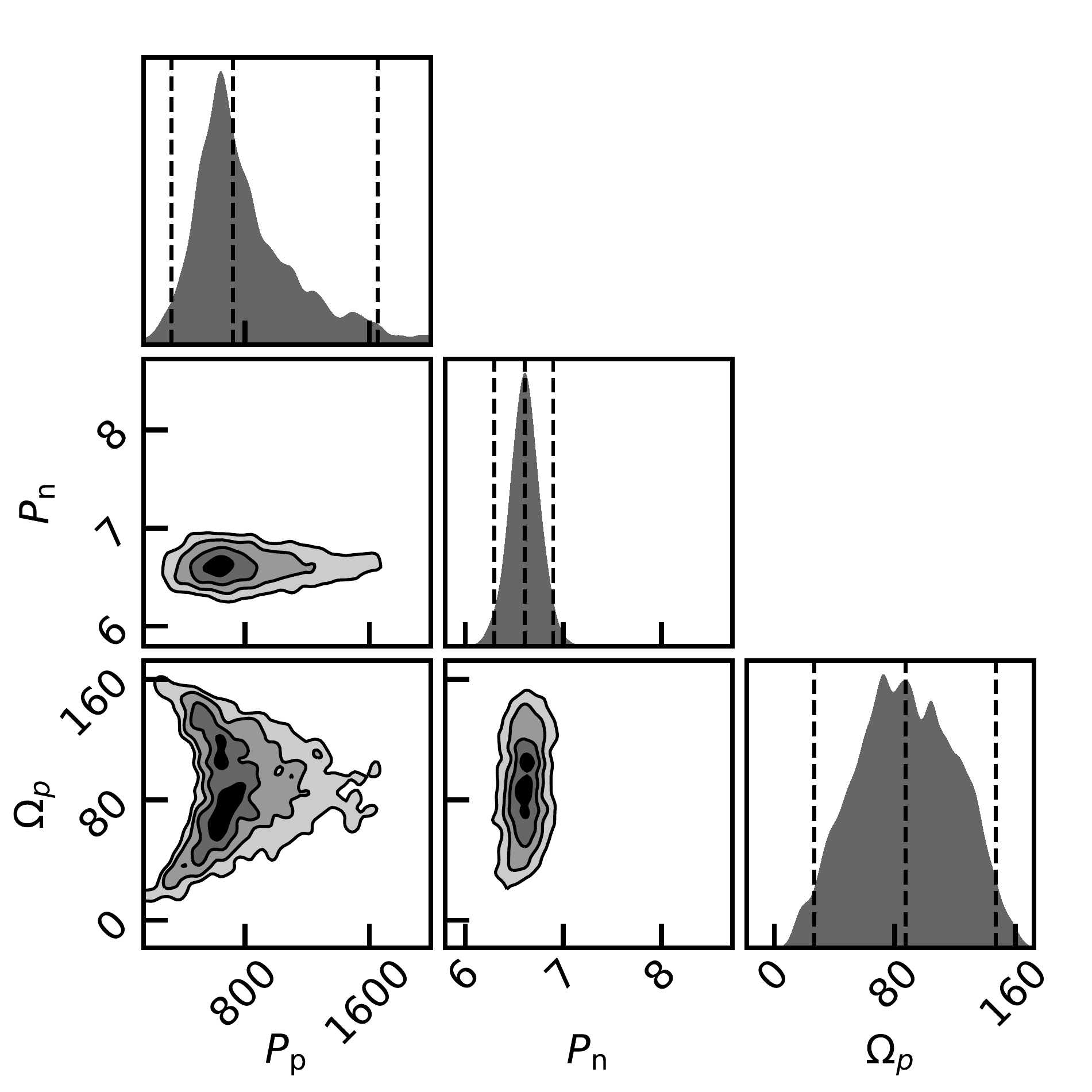}
    \caption{Posterior of the precession period $P_{\rm{p}}$, the nutation period $P_{\rm{n}}$, as well as the precession cone half-angle $\Omega_{\rm{p}}$.}
    \label{fig:corner_prec_nut}
\end{figure}

\cite{Britzen2018} analyzed the precession and nutation of the resolved jet components of OJ287. In the following section, we follow their nomenclature in order to derive the intrinsic time scales of the system from that observed (projected) precession and nutation. In \autoref{sec:sinus_model}, we modeled this precession as a sinusoidal modulation on top of a linear drift. In the context of a precession-nutation model, the linear trend describes (in first order of a Taylor-expansion) the precession and the additional sinusoidal modulation a nutation term. The model presented in \cite{Britzen2018}, which builds on model introduced by \cite{Abraham2000}, has seven parameters given in \autoref{tab:precession_nutation_model}.
While some of the older observations resolved the core structure, our prime observable is the position angle of the core on sky $\eta (t)$. Further, we do not detect a strong deviation from a linear trend of the core-position angle. Thus we cannot constrain the periodicity of the jet-precession $P_p$ directly, but can only constrain its value from the observed nutation and the linear trend. We fit the temporal evolution of the position angle using Equation 8 of \cite{Britzen2018}, and refer the reader to \autoref{sec:appendix-precession-nutation} for the mathematical details. We determine the posterior parameters of the model using \textit{dynesty} \citep{Skilling2006,Feroz2009,Skilling2004,Speagle2020}. \autoref{fig:corner_prec_nut} shows the posterior distributions of the precession period $P_{\rm{p}}$, the nutation period $P_{\rm{n}}$, and the precession cone half-angle $\Omega_{\rm{p}}$. As argued before, the nutation is driving the sinusoidal modulation of the position angle with a period $P_{\rm{n}}\sim 7$ years, while the linear trend is caused by a large scale precession of the jet with a largely unconstrained period (i.e., more than $200$ years, and less than $1800$ years), and the precession cone angle is constrained to be positive. 

\begin{table}[]
    \centering
    \begin{tabular}{l|c|r}
         Name & Abbreviation & Value\\
         \hline
         \hline
         Precession period & $\mathrm{P_p}$  & $(807\pm335)~\mathrm{yr}$\\
         Nutation period & $\mathrm{P_n}$ &  $(6.6\pm0.1)~\mathrm{yr}$\\
         Precession cone half-angle& $\mathrm{\Omega_{p}}$  &$(87\pm32)\degree$ \\
         Nutation cone half-angle& $\mathrm{\Omega_{n}}$  &$(\pm5.3\pm1.6)\degree$ \\
         Line-of-sight angle& $\mathrm{\Phi_0}$ &  $(12\pm39)\degree$\\
         Projected jet-cone-axis & $\mathrm{\eta_0}$  & $(62\pm1)\degree$\\
         Reference time & $\mathrm{t_0}$ & $(1994.8\pm0.2)~\mathrm{yr}$\\
         \hline
    \end{tabular}
    \caption{Model parameters and derived parameters of the precession-nutation model, as originally presented in \cite{Abraham2000} and \cite{Britzen2018}.}
    \label{tab:precession_nutation_model}
\end{table}

\section{Astrophysical origin of precession}
Jet precession is observed for many jetted AGNs, and their origin is typically attributed to either being of purely stochastic nature (i.e., without true periodicity), induced by disk-instabilities in tilted accretion disks or to be induced gravitationally by a pace-maker companion. 
Following the arguments presented by \cite{Vaughan2016} we cannot rule out a stochastic nature of the observed modulation in the M81* core region position angle. Nevertheless, so far, the apparently sinusoidal modulation of the precession angle proposed by \cite{Marti-Vidal2011} has both withstood the extrapolation to the future (albeit with a slightly increased linear slope), as well as the extrapolation to different frequencies (precession in the X band leading the C band). The latter is, however, not a proof of periodicity, as one would expect a frequency dependent-lag of the position angle also for a purely stochastic jet modulation.

Assuming that the observed modulation is truly periodic, two scenarios for its astrophysical origin may be of importance. For an accretion disc which is not aligned with the black hole's spin axis Lense--Thirring precession \citep{Thirring1918} induces a nodal precession of test particle orbits. The strength of this is effect is frequency dependent, and thus results in a precessing warp of the accretion disk. \cite{Fragile2007} demonstrated that this may lead to a constant period precession of the disk around the black hole. Such a disk-precession is thought to be the cause for Type-C Quasi-Periodic-Oscillations (QPOs) observed in X-ray binaries \citep[e.g.,][]{Stella1998}, and can be used to estimate the black hole mass and spin under certain assumptions \citep[e.g.,][]{Ingram2014}. Building on the simulations by \cite{Fragile2007}, \cite{Liska2018} demonstrated that titled accretion flows are able to launch jets, and that the jet precession is aligned with precession of the disk. In this set of simulations the amplitude of the jet precession depends on the separation from the black hole, dropping from $100\degree$ at a separation of $R\sim 10 R_g$ to $\sim 50\degree$ at $100 R_g$. Further, \cite{Fragile2007} found an empirical relation between the precession period of the disk: $T_{p}\sim 0.3(m/M_{\sun})~s$, which corresponds to roughly $0.2$ years assuming a black hole mass of $2 \times 10^7M_{\sun}$. Both the amplitude of the oscillation as well as the precession period are in slight tension with the observed values ($2\Theta_p\sim 7\degree$, $P_p\sim7\mathrm{yr}$). Assuming a precession-nutation model, the amplitude of precession is more similar to the expected value ($\sim80\degree$), however the much longer period in this case is even harder to contextualize. In this simple argumentation we have, however, not accounted for projection effects, and none of the simulations have been tailored to M81* (i.e. unknown spin, and accretion-disk tilt), and have ignored that the disk precession timescales dependent on the initial disk mass \cite{Liu2002}. We therefore suggest that the Lense-Thirring precession scenario may well be applicable in the case of M81*. 

Finally, a gravitational pacemaker may explain the observed precession of the M81* position angle. Such a scenario has been found in the binary system SS433 \citep[e.g.,][]{Stephenson1977,Clark1978} where two large scale precessing jets create a corkscrew like structure. For this system, the precession is thought to originate from gravitational torque of the donor star on the accretion disk of a compact object which is either a black hole or a neutron star \citep[slaved disk model,][]{Roberts1974, Waisberg2019}. A similar scenario has been proposed for OJ 287 and 3C 345, where an orbital modulation is present both in the light curve as well as the jet components \citep{Lobanov2005, Britzen2018}. Further \cite{Caproni2013} found a $12.1~\mathrm{yr}$ precession period in BL Lacertae. \cite{Britzen2018} found a precession period of $\sim23~\mathrm{yr}$, on top a much faster $\sim 1~\mathrm{yr}$ nutation period and derived a binary separation between $0.001~\mathrm{pc}<d<0.1~\mathrm{pc}$. For 3C 345, \cite{Lobanov2005} found a precession period of $\sim 10~\mathrm{yr}$ of the position angle, and which exhibited a linear trend, which is very similar to the behavior of M81* (short $\sim 7 ~\mathrm{yr}$ nutation period, a linear trend, which we interpret as a long-period precession of $\sim 800~\mathrm{yr}$).

\noindent If we interpret the seven year period inferred as the orbiting period of the companion object, we can derive the binary semi-major axis as well as the gravitational binary merger time. An orbiting secondary black hole induces gravitational torques on the accretion disc, which results in the precessing motion in the opposite sense to the disc rotation. The precessing motion is accompanied by the short-term nutation motion caused by the torque of a similar magnitude. However, the amplitude is smaller than for the precession by the ratio of the precession and the orbital frequencies. Following \cite{Katz1982}, see also \cite{Caproni2013}, the nutation angular frequency is equal to twice the difference of orbital and precession angular frequencies,

\begin{equation}
    \omega_n = 2(\omega_{\rm{orb}} - \omega_{p}), \label{eqn:orbit}
\end{equation} 

\noindent where $\omega_p$ is negative because of the opposite sense with respect to the orbital motion. Using \autoref{eqn:orbit}, the orbital period can be expressed as

\begin{equation}
    P_{\rm{orb}} = \dfrac{2P_n}{1 - 2P_n/P_p} \approx 2P_n . \label{eqn:period}
\end{equation} 

\noindent The putative SMBHB in M81 has an orbital period of $P_{\rm{orb}} \sim 14$ to $15$ years for $P_p \geq P_n$, which seems to be the case based on the long-term linear drift. The semi-major axis of the binary system can be estimated from the third Kepler law

 \begin{equation}
     a_{\rm{bin}} \approx 1573 \left( \dfrac{M_{\rm{tot}}}{2\times 10^7M_\odot} \right)^{1/3} \left( \dfrac{P_{\rm{orb}}}{14~\mathrm{yr}} \right)^{2/3} \mathrm{AU}, \label{eqn:axis}
 \end{equation}

\noindent which corresponds to $a_{\rm{bin}} \sim 8000 R_g$ ($R_g = GM_{\rm{tot}}/c^2$) in gravitational radii if the primary SMBH dominates the total mass. Using the primary mass fraction of $x_p = m_1/M_{\rm{tot}} \sim 0.9$ (the secondary mass fraction is $x_s \sim 0.1$), the binary merger timescale is

\begin{align}
    \tau_{\rm{merge}} &= 5/256 \dfrac{c^5}{G^3} \dfrac{a_{\rm{bin}}^4}{x_p x_s M_{\rm{tot}}^3} \\
    &= 2.77\left( \dfrac{a_{\rm{bin}}}{1573 ~\mathrm{AU}}\right)^4  \left(\dfrac{x_p}{0.9}\right)^{-1} \left( \dfrac{x_s}{0.1}\right)^{-1} \left( \dfrac{M_{\rm{tot}}}{2\times 10^7 ~ \mathrm{M_\odot}}\right)^{-3} ~\mathrm{Gyr}.\label{eqn:merge}
\end{align}

\noindent For an even more extreme ratio, $x_p \sim 1$, $x_s \sim 10^{-2}$, we obtain $\tau_{\rm{merge}} \sim 25~\mathrm{Gyr}$, hence the system would generally be long-lived.

\section{Conclusions}
We present novel observations of the core region of M81*. Our observations are consistent with a precession of the jet of this galaxy, which was first proposed by \cite{Marti-Vidal2011}. The jet precession period is roughly 7 years. On top of the precession (amplitude $\sim7\degree$), the jet exhibits a small linear drift of roughly $0.5\mathrm{\degree/yr}$. However, we cannot confirm that the flare observed in the years 1998 to 2002 is connected to the precession of the jet for instance through Doppler boosting. Our sparse flux measurements do not show elevated flux through the years 2017 to 2019. Further, historic data does not show any flaring activity since the end of the last flare in 2002. We can rule out a self-similar modulation of the light curve with the period of the modulation. However, the sampling of the light curve is too spare to definitely rule out any modulation scenario. Thus we consider it more likely that the flare observed in the years from 1998 to 2002 was not caused by variable Doppler boosting. 
Our observations are consistent with either a Lense--Thirring induced precession of the jet, or binary-induced precession of the jet. It aligns itself with observations of other precessing jets such as in 3C345 and OJ287, which both show a fast nutation-like component as well as a slower, but larger amplitude precession-like variation of the jet position angle. If Lense--Thirring precession is responsible for the observed jet precession in all three AGN, the underlying coupling of the accretion disk precession to the jet precession would show a remarkable self-similar coupling through vastly different accretion regimes. While both 3C 345 and OJ 287 accrete close to their Eddington limit, M81* belongs to the class of radiatively inefficient accretion flows. This may hint that the jet-physics responsible for the observed precession in these systems maybe accretion-flow-rate and accretion-flow-state independent. 
On the other hand, if a binary black hole is responsible for the apparent precession, then it may be the closest super-massive or intermediate mass binary black hole candidate. 
However, more data is necessary to finally confirm the precessing nature of M81*, and in particular the proposed frequency dependent time lag. 

\begin{acknowledgements}

This publication is part of the M2FINDERS project which has received funding from the European Research Council (ERC) under the European Union's Horizon 2020 Research and Innovation Programme (grant agreement No 101018682). J.D. acknowledge support from NSF AST-2108201, NSF PHY-1903412 and NSF PHY-2206609. J.D. is supported by a Joint Columbia University/Flatiron Research Fellowship, research at the Flatiron Institute is supported by the Simons Foundation. MZ acknowledges the financial support by the GACR EXPRO grant no. GX21-13491X. 

\end{acknowledgements}

\bibliography{bib/bib_m81}{}
\bibliographystyle{aasjournal}
\appendix
\section{Best fit model including the newest observation}
In this appendix we show the best fit model detailed in the third column of \autoref{tab:model-fit}, as well as the linear drift model (fourth column of \autoref{tab:model-fit}) defined as:
\begin{equation}
    \theta(t) = \beta\cdot (t- t_0) \label{eqn:lin_model}.
\end{equation}
\begin{figure}
    \centering
    \includegraphics[width=0.485\textwidth]{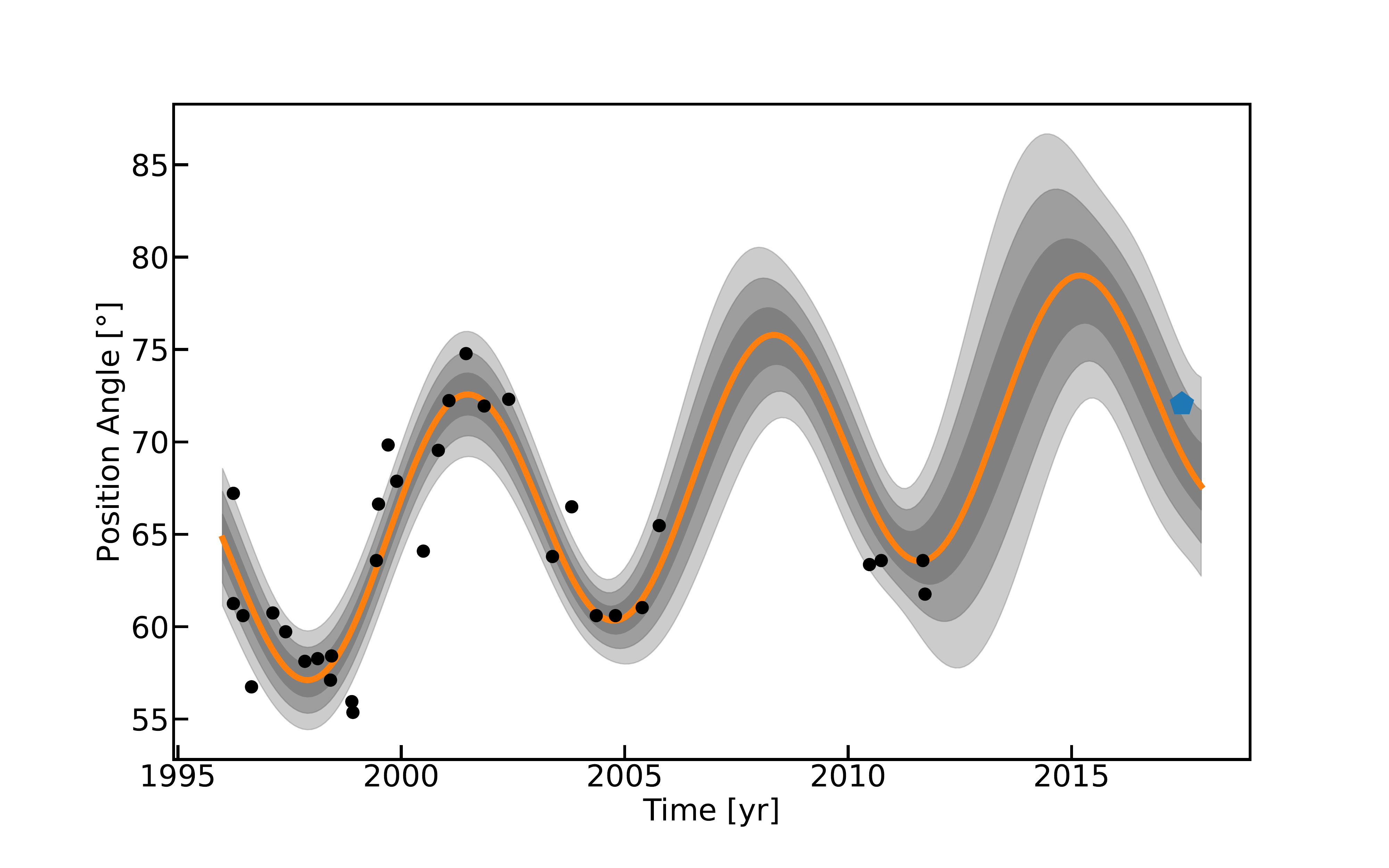}
    \caption{Same as \autoref{fig:fit-2011}, but including the 2017 datum in the fit.}
    \label{fig:fit_2017}
\end{figure}
\begin{figure}
    \centering
    \includegraphics[width=0.485\textwidth]{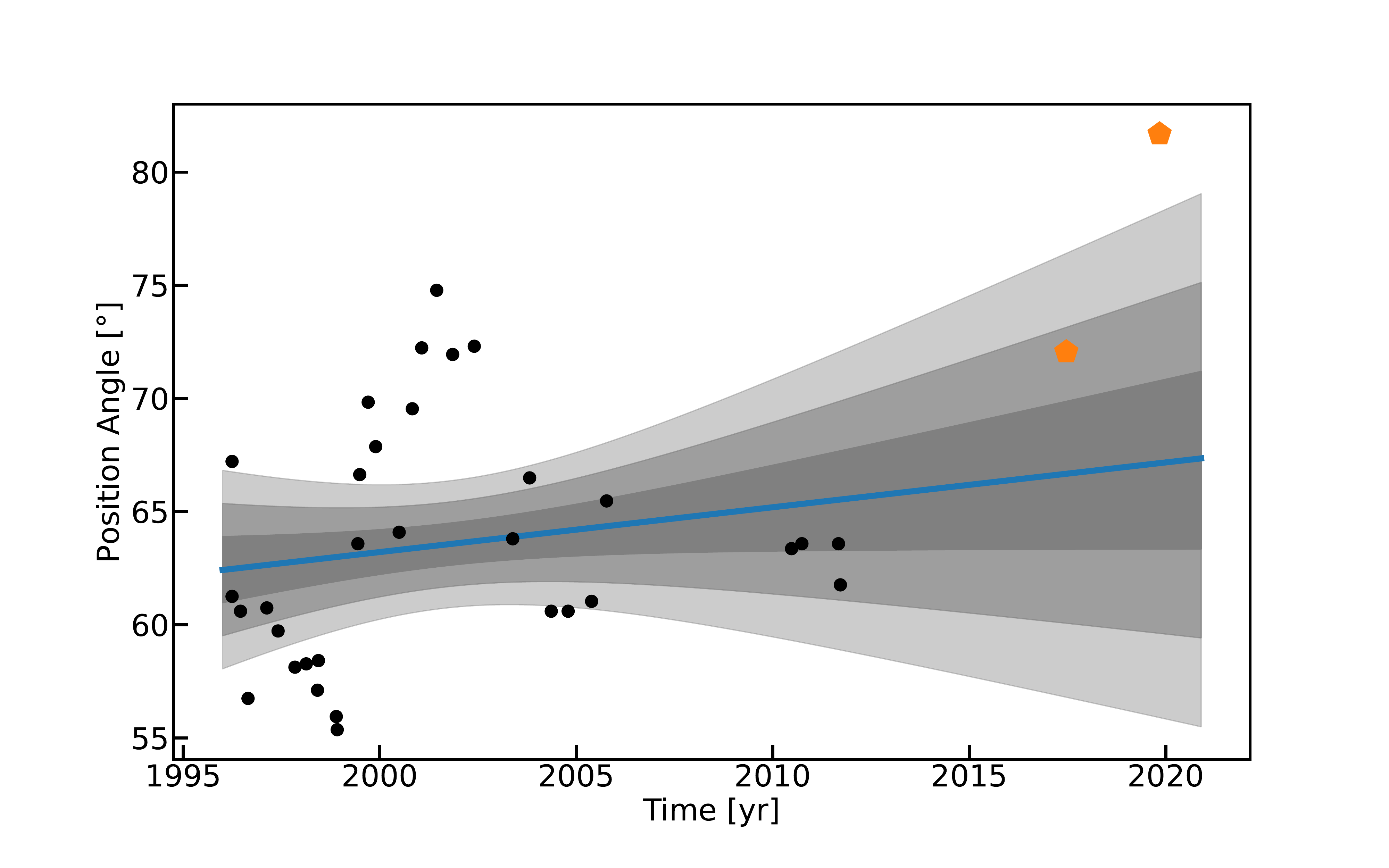}
    \caption{Same as \autoref{fig:fit-2011}, but for a linear drift model and including the 2017 and 2019 data in the fit.}
    \label{fig:linear_mod}
\end{figure}

\section{Frequency-dependet shift excluding the 2019 observations}
\begin{figure}
    \centering
    \includegraphics[width=0.485\textwidth]{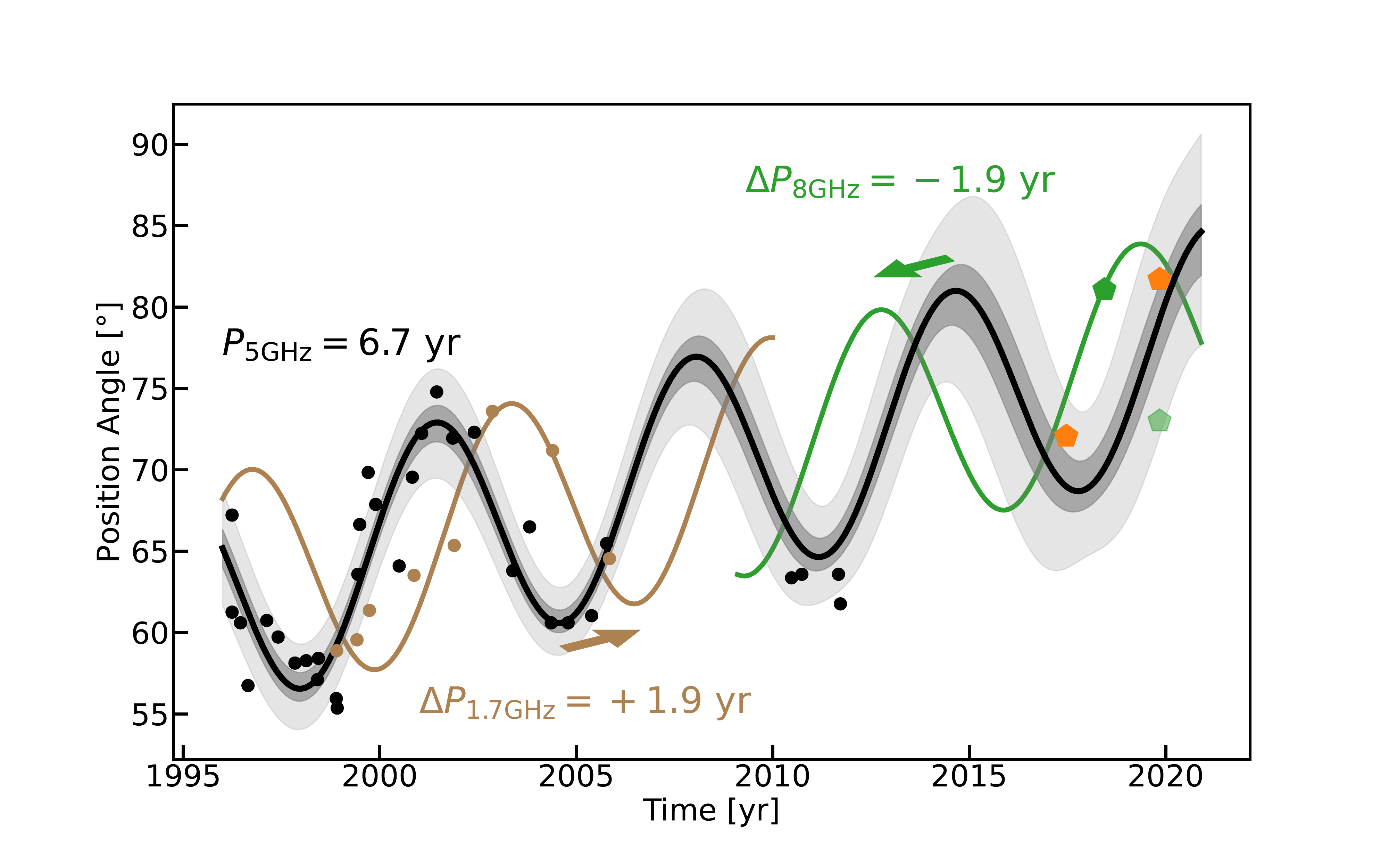}
    \caption{Same as \autoref{fig:frequency_dependet}, but with a -1.9 year shift of the 8 GHz model instead of -3.5 years.}
    \label{fig:shift_1.9years}
\end{figure}

We show the same Figure as \autoref{fig:frequency_dependet} in \autoref{fig:shift_1.9years}, but with a model shifted by $-1.9$ years instead of $-3.5$ years. This is consistent with the well determined 2018 8 GHz observation, but inconsistent with the 2019 observations.

\section{Full posterior of the precession nutation model}\label{sec:appendix-precession-nutation}
\begin{figure*}
    \centering
    \includegraphics[width=0.985\textwidth]{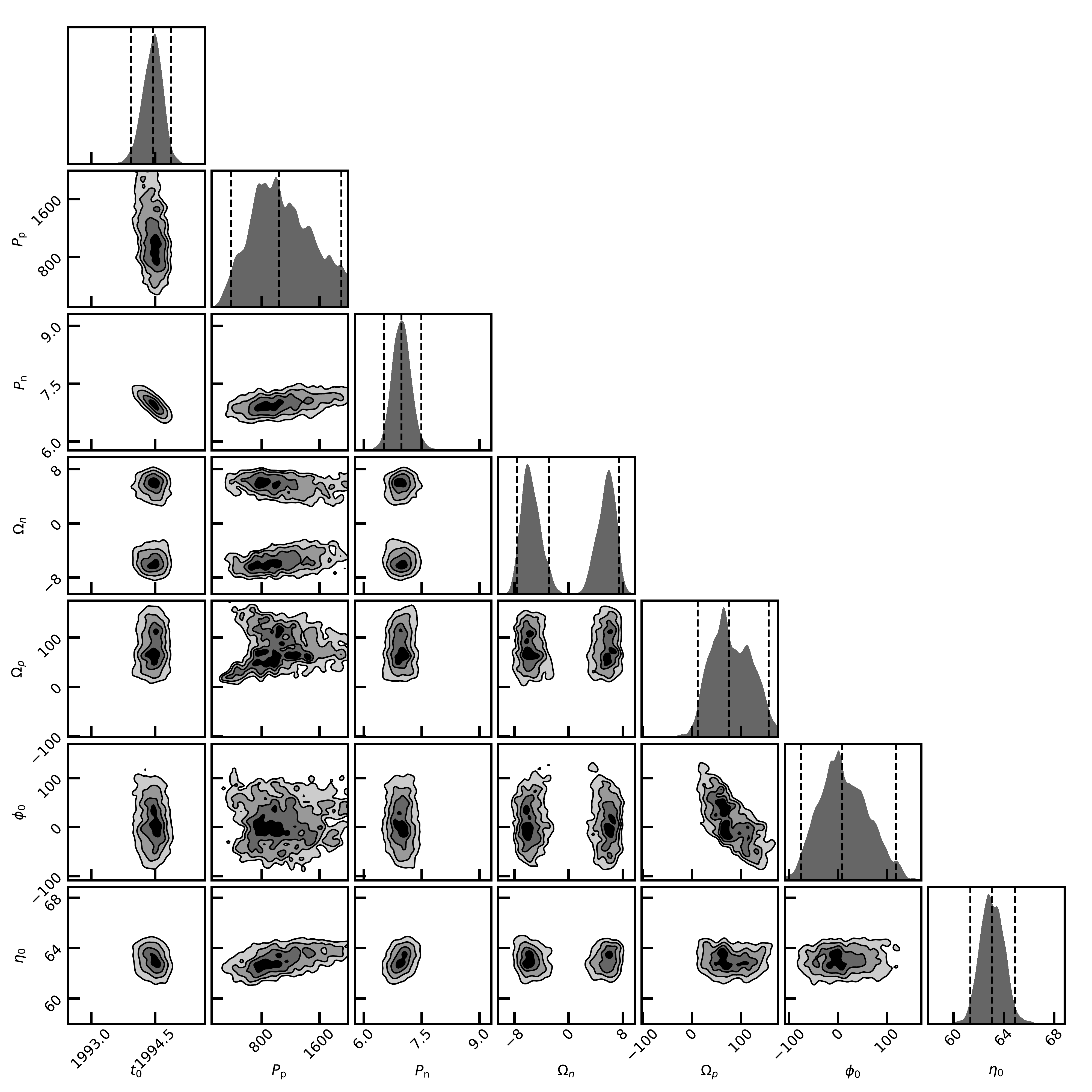}
    \caption{Joint posterior of the precession nutation model.}
    \label{fig:precession_nutation_model_full}
\end{figure*}

The full posterior of the precession nutation model is given in \autoref{fig:precession_nutation_model_full}

\section{Maps of observation}\label{appendix:models}
Below we plot the maps, best-fit models of the respective C- and X-band observations. \autoref{fig:c-band} shows the C-band data, \autoref{fig:x-band} shows the X-band data. \autoref{tab:obsdetails} gives an overview of the observation details.
\begin{table*}[]
    \centering
    \begin{tabular}{l|c|c|c|c}
    Obs. date & Band & Beam size [mas] & Beam angle [$\degree$] & Clean residual RMS [Jy/beam] \\
    \hline
    2017-06-21 & C & $0.902\times 0.86$ & $-65.3$ & $0.0004$\\
    2018-06-10 & X & $0.911\times 0.594$ & $-12$ &  $0.0002$\\
    2019-11-04 & C & $1.78 \times 1.2$ & $-30.3$ & $0.0003$\\
    2019-11-04 & X & $1.01 \times 0.755$ & $-29.1$ & $0.0003$\\
    \end{tabular}
    \caption{Achieved resolution and beam configuration for the four observations.}
    \label{tab:obsdetails}
\end{table*}
\begin{figure}
\centering
\begin{subfigure}{.5\textwidth}
  \centering
  \includegraphics[width=.55\linewidth, angle=0, trim={1.2cm 2.cm 0.5cm 3.6cm},clip]{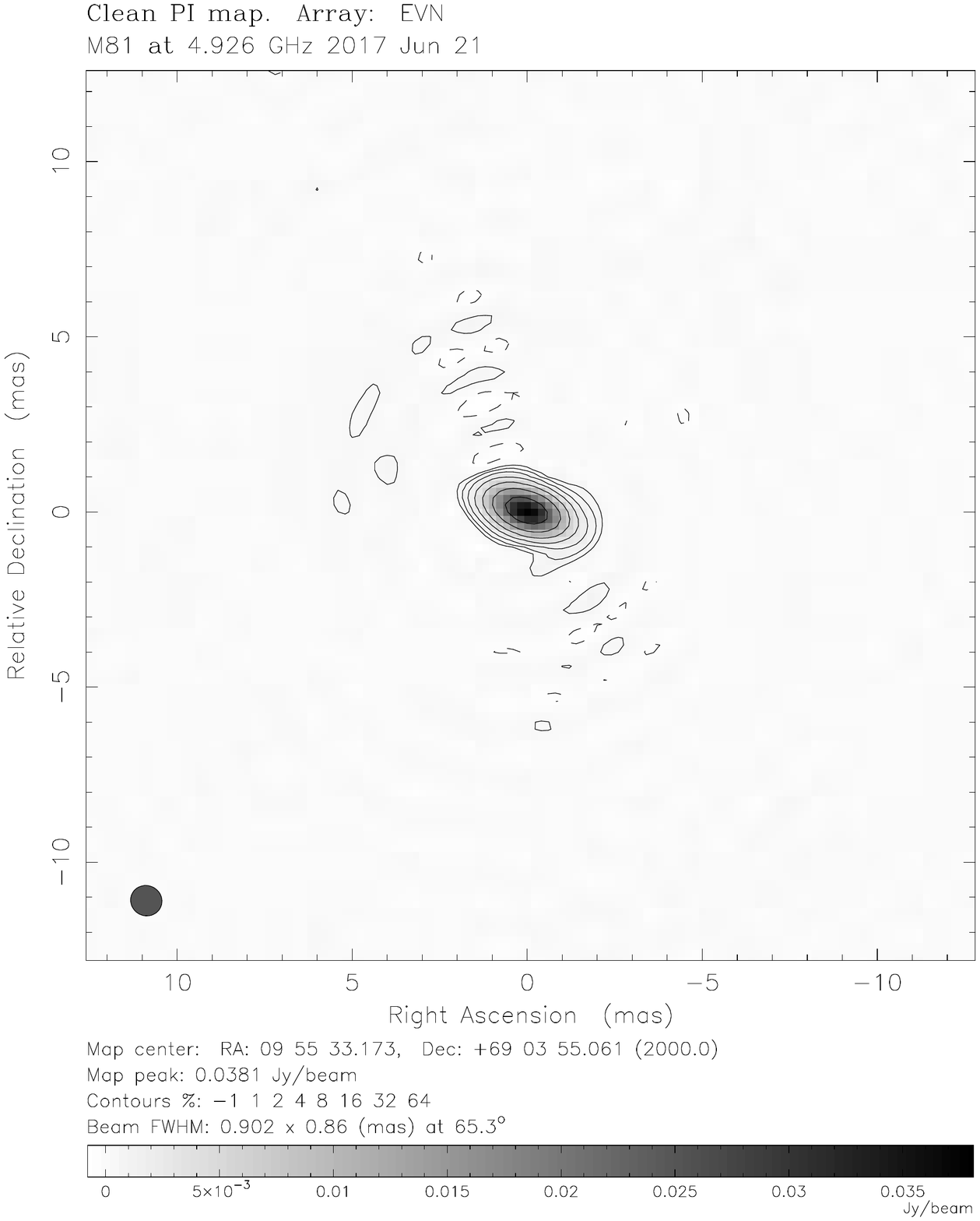}
  \label{fig:sub1}
\end{subfigure}%
\begin{subfigure}{.5\textwidth}
  \centering
  \includegraphics[width=.8\linewidth, angle=270]{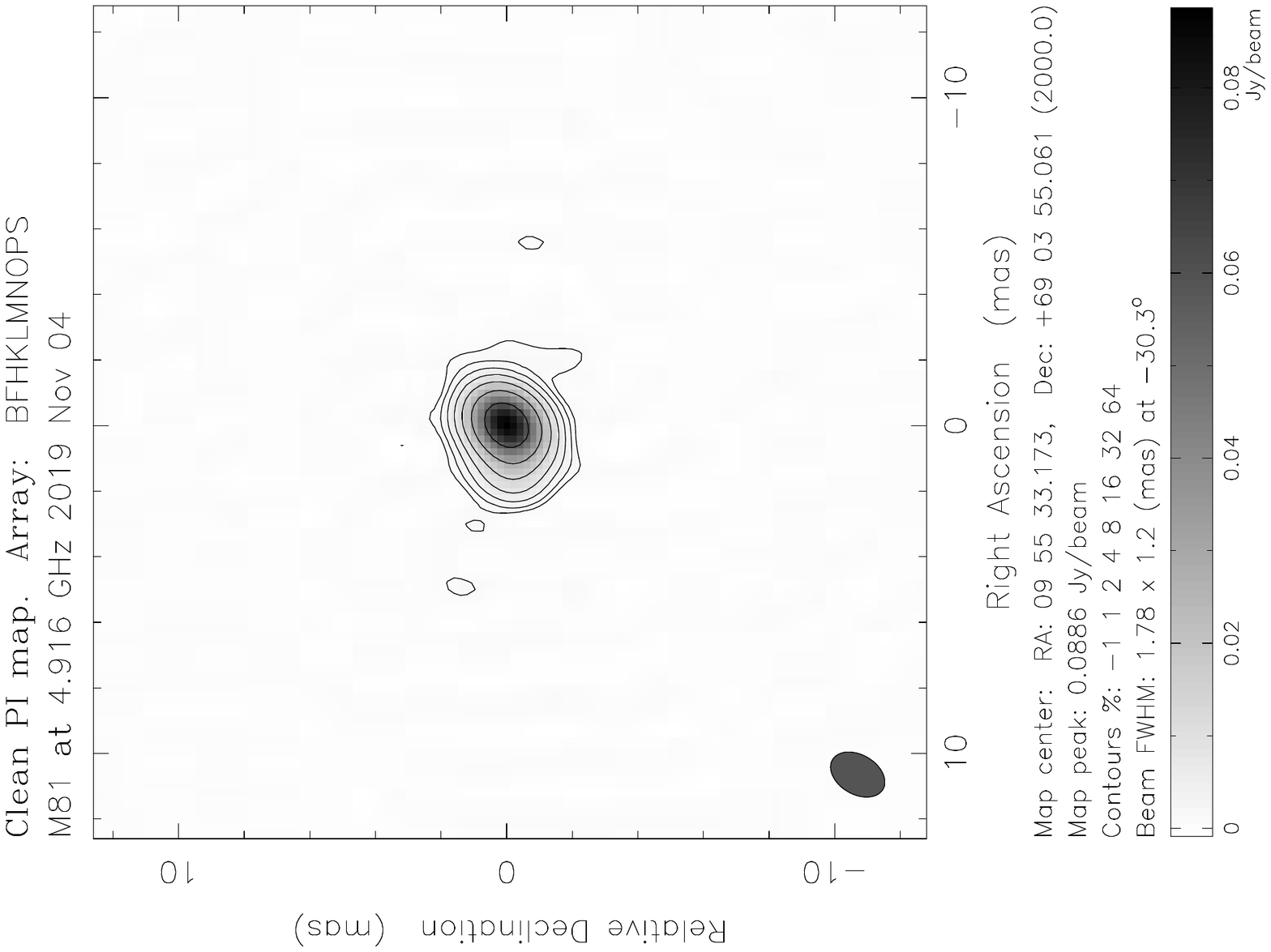}
  \label{fig:sub2}
\end{subfigure}
\caption{Maps of the observations obtained in C-band. The beam size is indicated by the grey ellipse. Sub-figure (a) shows the C-band observation obtained with the EVN on June, 2017; sub-figure (b) shows the C-band observation obtained with the VLBA on November, 2019..}
\label{fig:fig:c-band}
\end{figure}

\begin{figure}
\centering
\begin{subfigure}{.5\textwidth}
  \centering
  \includegraphics[width=0.79\linewidth, angle=270]{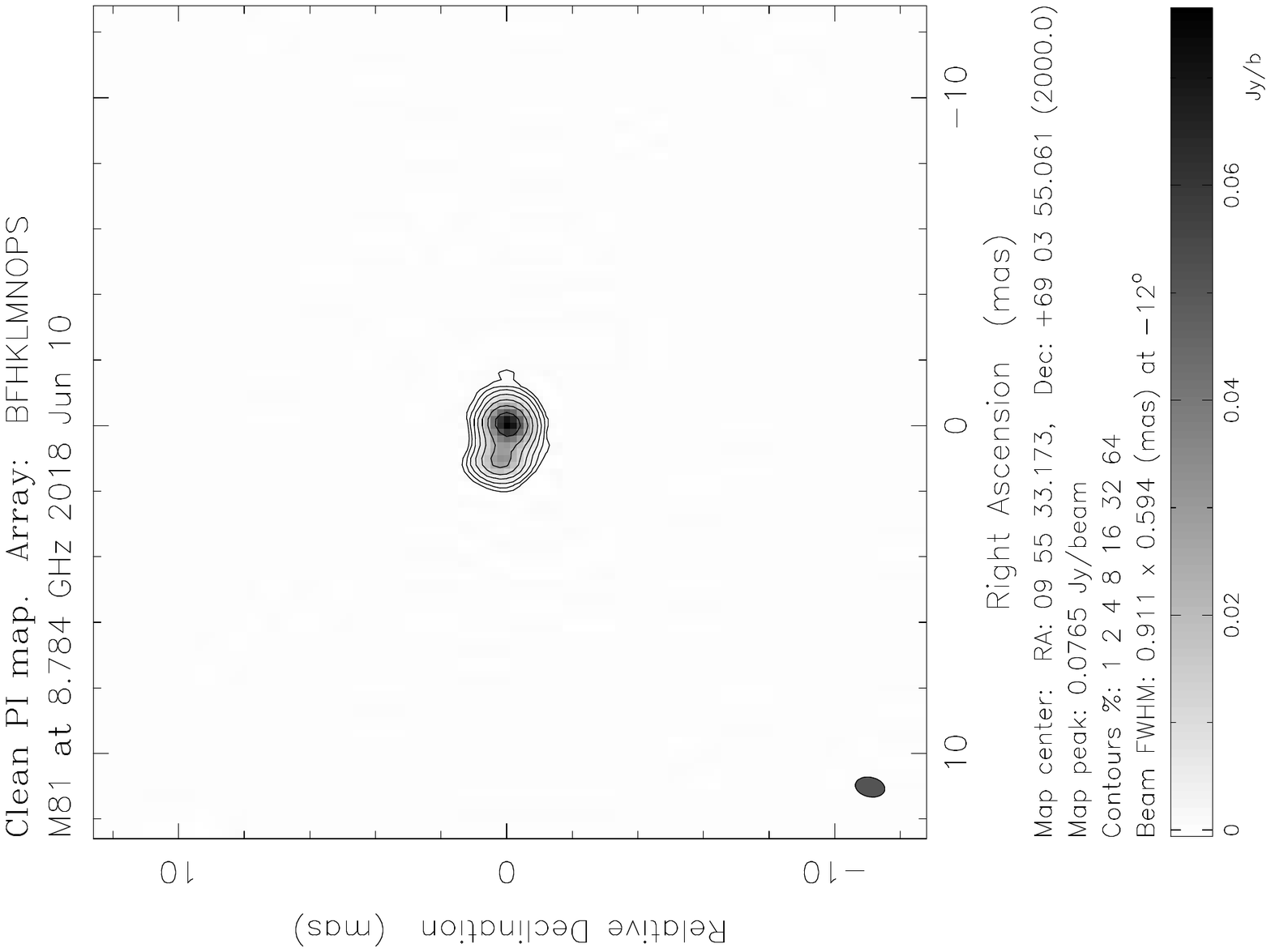}
  \label{fig:sub3}
\end{subfigure}%
\begin{subfigure}{.5\textwidth}
  \centering
  \includegraphics[width=0.8\linewidth, angle=270]{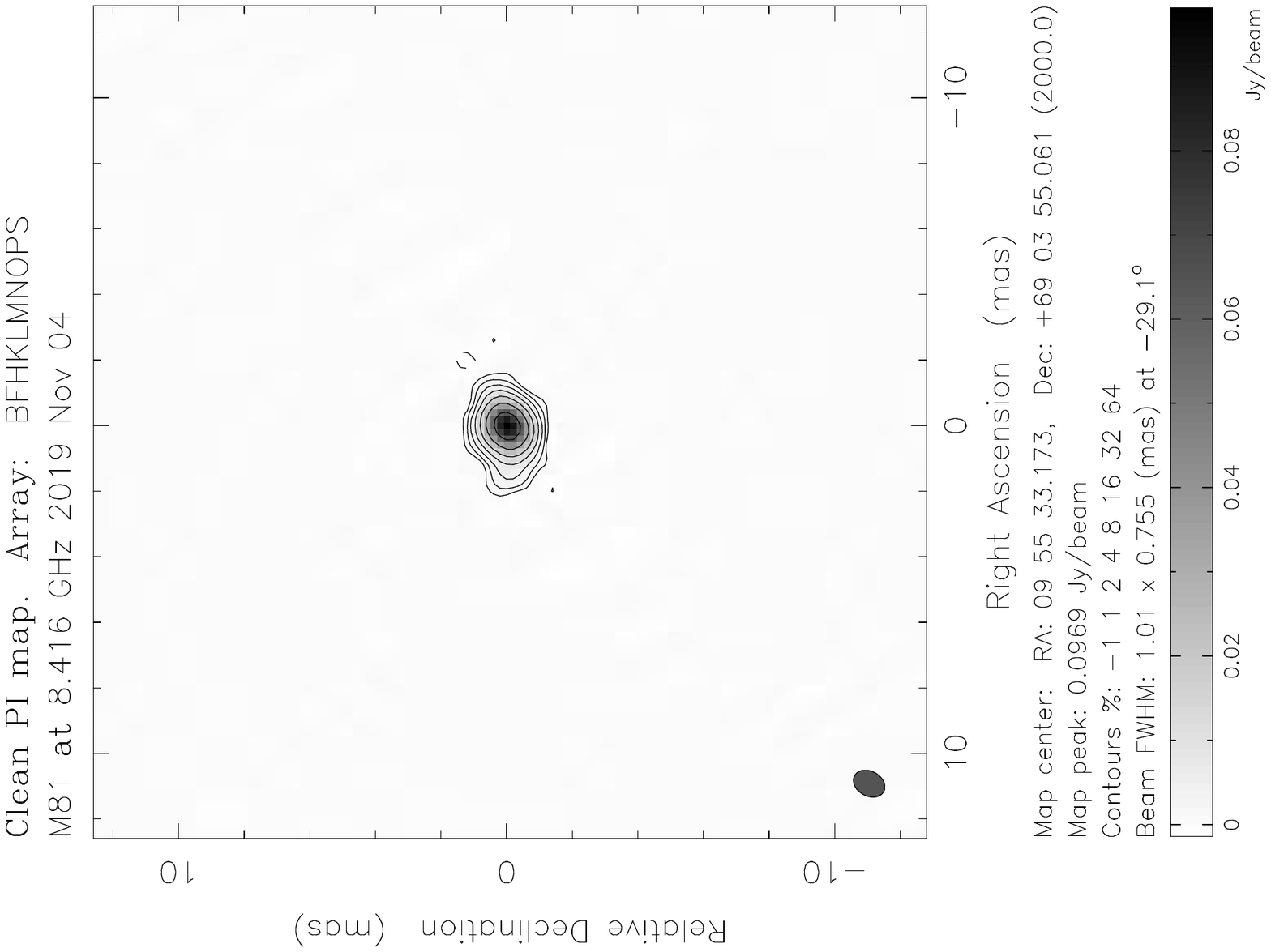}
  \label{fig:sub4}
\end{subfigure}
\caption{Same as \autoref{fig:fig:c-band}, for observations obtained in X-band. The beam size is indicated by the grey ellipse. Sub-figure (a) shows the X-band observation obtained with the EVN on June, 2018; sub-figure (b) shows the X-band observation obtained with the VLBA on November, 2019.}
\label{fig:x-band}
\end{figure}

\section{Participating stations}
\autoref{tab:participating_stations} gives the participating stations and their abbreviations.
\begin{table*}[]
    \centering
    \begin{tabular}{c|c|c}
         abbreviation & Observatory name & Location\\
         \hline
         \hline
         EF & Effelsberg & Germany\\
         IR & Irbene & Latvia \\
         YS & Yonsei & R. o. Korea\\
         TR & Torun & Poland\\
         SH & Shanghai Astronomical Observatory& P.R. China\\
         MC & Medicina & Italy\\
         WB & Westerbork Synthesis Radio Telescope & the Netherlands\\
         JB & Jodrell Bank Observatory & United Kingdom\\
         BR & Brewster & USA\\
         FD & Fort Davis & USA\\
         HN & Hancock & USA\\
         KP & Kitt Peak & USA\\
         LA & Los Alamos & USA\\
         MK & Mauna Kea & USA\\
         NL & North liberty & USA\\
         OV & Owens Valley & USA\\
         PT & Pie Town & USA\\
         SC & St. Croix & USA\\
         \hline
    \end{tabular}
    \caption{Station names, abbreviations and location of the participating EVN observatories.}
    \label{tab:participating_stations}
\end{table*}

\end{document}